\documentclass[a4paper,12pt]{article}

\usepackage[english]{babel}
\usepackage{amsmath}
\usepackage{amsfonts}
\usepackage{amssymb}
\usepackage{a4wide}
\usepackage[pdftex]{graphicx}
\usepackage{graphicx}
\usepackage{cite}

\renewcommand{\not}[1]{#1 \hskip-0.475em /}

\begin{document}
\thispagestyle{empty}

\begin{flushright}
{\small
TUM-HEP-848/12\\
TTK-12-31\\
July 20, 2012}
\end{flushright}

\vskip1.5cm
\begin{center}
{\Large\bf\boldmath
Soft-collinear gravity}
\vspace{\baselineskip}

\vspace{1cm}
{\sc M.~Beneke}${}^{a,b}$
and
{\sc G.~Kirilin}${}^{a,b}$\\[5mm]
${}^a${\it Physik Department T31,\\
Technische Universit\"at M\"unchen,\\
James-Franck-Stra\ss e~1, D - 85748 Garching, Germany\\[0.3cm]
${}^b$
{\it Institut f{\"u}r Theoretische Teilchenphysik
 und Kosmologie,\\ RWTH Aachen University,}\\
{\it D--52056 Aachen, Germany}
}\\[0.5cm]

\vspace*{1cm}
\textbf{Abstract}\\
\vspace{1\baselineskip}
\parbox{0.9\textwidth}{
We study collinear and soft singularities in perturbative quantum
gravity by constructing an effective field theory similar to
soft-collinear effective theory for QCD (SCET). We find that the
soft sector exhibits factorization properties similar to those of SCET.
The collinear sector is, however, quite different. While the leading-power
collinear effective Lagrangian is trivial, the presence of the metric
field $h_{++}$ with negative scaling dimension allows for collinear
divergences in loop diagrams with couplings to non-collinear sources.
We provide a compact proof of the
well-known fact that there are no collinear singularities in
perturbative quantum gravity by demonstrating the decoupling
of $h_{++}$ from the sources. We briefly discuss the connection of our
approach to recent work by Akhoury et al. (Phys. Rev.~D84 (2011) 104040)
as well as to the Weinberg's original paper (Phys. Rev.~140 (1965) B516),
where the cancellation of the collinear singularities was demonstrated
for the first time in the eikonal approximation.}
\end{center}

\newpage
\setcounter{page}{1}


\newpage

\subsection*{1 Introduction}

The story of collinear divergences in general relativity is rather short. In
Ref.~\cite{Weinberg1965}, using the eikonal approximation, Weinberg
shows that no additional divergences in gravitational radiation appear in the
limit of massless colliding particles. Utilizing the gravitational Ward
identity, Akhoury et al.~\cite{Akhoury2011} demonstrate without reference
to the eikonal approximation that the collinear
singularities cancel to any perturbative order when all relevant
diagrams are summed over.

There is a simple qualitative explanation of the absence of singular
collinear graviton radiation. Consider an energetic particle with
virtuality much less than its three-momentum squared $\mathbf{p}^{2}$,
emitting a nearly collinear graviton with momentum
$\mathbf{k}$ such that the angle $\theta$ between $\mathbf{p}$ and
$\mathbf{k}$ is small. The near mass-shell singularity of the emitting
particle propagator yields a factor $\theta^{-2}$ for the splitting
amplitude. The on-shell graviton is produced in a state with the definite
helicity $\pm 2$.Due to helicity and angular momentum conservation, the
splitting amplitude should be proportional to the component of the graviton
wave function with vanishing projection of angular momentum along the
momentum of the initial particle. Quantizing the radiation field in
the spherical basis with single-particle states
$|\mathbf{k} j m;\lambda\rangle$,
where $\lambda$ denotes helicity and $jm$ the angular momentum quantum
numbers with respect to the quantization axis $\mathbf{p}$, this implies
that the emitted graviton must be in a state
$|\mathbf{k} j \,0;\pm 2 \rangle$. The angular dependence of this
state is given by a spin-weighted spherical harmonic or Wigner function
$D_{\pm2,0}^{j}\left(  \mathbf{k}\right)\propto \sin^2\theta$,
which tends to zero as $\theta^{2}$ in the $\theta\rightarrow0$ limit.
Thus, the splitting amplitude has no singularity in the collinear limit. In
contrast to the graviton, a massless vector boson contributes
as $D_{\pm1,0}^{j}\left(  \mathbf{k}\right)  \sim\theta$, which leads to
a $\theta^{-1}$ singularity in the amplitude and the well-known
logarithmic divergence $\mathrm{d}\theta/\theta$ in the differential
cross section.

The above argument refers to physical polarization states of the graviton
and thus does not cover the properties of individual Feynman amplitudes
in general, in particular covariant gauges, which do have collinear
divergences. In order to demonstrate the absence of collinear
singularities in a physical process, one needs a factorization theorem
that controls the collinear interactions of the unphysical graviton
modes and their coupling to a non-collinear environment (``source'').
In Ref.~\cite{Akhoury2011} the gravitational Ward identity is
employed to provide a diagrammatic proof of the factorization and
cancellation of collinear divergences, extending Weinberg's analysis
beyond the eikonal approximation.

In order to single out the singular diagrams, the authors of
Ref.~\cite{Akhoury2011} use power-counting rules which, as mentioned
in this paper, are very similar to those used to construct the soft-collinear
effective theory (SCET) for
QCD \cite{Bauer2001,Bauer2002,Beneke2002,Beneke2003}. In addition to
power counting, SCET often simplifies the algebra of factorization
proofs, since it displays the relevant properties in the Lagrangian,
and avoids reference to individual diagrams. This motivates us to
reconsider the problem of soft and collinear graviton physics by
constructing ``soft-collinear gravity''. That is, following the lines of
SCET,  we analyze the coupling of soft and collinear field degrees of
freedom at the level of an effective Lagrangian instead of classifying
all relevant Feynman graphs. We also find it interesting to compare
soft-collinear gravity to the SCET for gauge fields, which reveals
similarities and differences. As will be shown below, one of the
differences is the presence of a metric field component with negative
scaling dimension, which complicates the correspondence between the power
expansion of the Lagrangian and the scaling of diagrams. The negative-scaling
component $h_{++}$ also plays the crucial role in the interaction of
collinear modes with the non-collinear environment; controlling the
interactions of $h_{++}$ is the essence of factorization. The field
$h_{++}$ could be eliminated by choosing a non-covariant physical gauge,
as is expected from the arguments above. However, our aim is
to demonstrate the factorization in a covariant gauge, which will be
accomplished by a universal field redefinition. For the sake of
completeness, we also consider the interactions of soft gravitons,
which share many similarities with soft gauge fields.

\subsection*{2 Power-counting rules}

The version of SCET we use as a template is based on the position-space
representation~\cite{Beneke2002, Beneke2003}.
To refrain from repetition, we review here only the key ideas of the
effective theory construction, and refer the reader to Ref.~\cite{Beneke2002}
for further information. The theory employs separate fields for the
collinear and the soft modes with small virtuality compared to the
large scale of the process. Each collinear region is characterized by
a certain light-like four-vector $n_{-}^\mu$. The complement light-like
vector is denoted $n_{+}^\mu$, such that $n_{+}\cdot n_{-}=2$, and it is
convenient to introduce the notation
\begin{equation}
p\_=n_{-}\cdot p,\quad p_{+}=n_{+}\cdot p
\end{equation}
for the light-cone components.
A collinear four-vector $p$ is assumed to have the following
scaling of its components:
\begin{equation}
p_{+}\sim Q,\quad p_{-}\sim\lambda^{2}Q,\quad
p_{\perp}^\mu=p^\mu-\frac{p_{+}n_{-}^\mu+p_{-}n_{+}^\mu}{2}\sim\lambda Q,
\label{eq:ColScale}
\end{equation}
where $Q$ is the hard scale in a process and $\lambda$ is a small
dimensionless parameter. A soft four-vector $q$ has scaling behavior
\begin{equation}
q^\mu\sim \lambda^2 Q
\label{eq:SoftScale}
\end{equation}
for any of its components. The effective fields are defined to
create or destroy particles with a certain
momentum scaling. Scaling rules for field components can be extracted from the
field two-point correlators. For example, a collinear fermion field
$\psi_{c}$ is decomposed into two components with different scaling (see
Ref.~\cite{Beneke2002}):
\begin{equation}
\psi_{c}=\xi+\zeta,\quad\xi=\frac{\not n  _{-}\not n  _{+}}{4}\psi_{c}%
\sim\lambda,\quad\zeta=\frac{\not n  _{+}\not n  _{-}}{4}\psi_{c}\sim
\lambda^{2}, \label{eq:PsiScale}
\end{equation}
while the components of the collinear gauge field $A^\mu_c$ scale like a
collinear momentum. The small field component $\zeta$ is integrated
out and not part of the soft-collinear Lagrangian.

The same procedure can be applied to the gravitational field
$h_{\mu\nu}$, defined as the metric deviation from flat space,
\begin{equation}
g_{\mu\nu}=\eta_{\mu\nu}+h_{\mu\nu}.
\label{eq:weakG}
\end{equation}
The corresponding expansion of the Einstein-Hilbert
action reads as follows:
\begin{eqnarray}
S  &  = &\frac{1}{16\pi G_N} \int\mathrm{d}^{4}x\,\sqrt{-g} \,R
\nonumber\\
&=& \frac{1}{2\kappa^2}
\int\mathrm{d}^{4}x\,\left[  \,\partial_{\alpha}%
h_{\mu\nu}\partial^{\alpha}h^{\mu\nu}-\,\partial_{\alpha}h\,\partial^{\alpha
}h-2\,\partial_{\mu}h^{\mu\nu}\left(  \partial_{\alpha}h_{\nu}^{\alpha
}-\partial_{\nu}h\right)  +O\!\left(  h^{3}\right)  \right],
\label{eq:Einst}
\end{eqnarray}
where $h=h_{\alpha}^{\alpha}$, $\kappa=\sqrt{32\pi G_N}$, and $G_N$ is the
Newton constant. After expansion in $h_{\mu\nu}$ indices are raised
and lowered with the flat-space metric. The gravity coupling
$\kappa$ is dimensional; the dimensionless parameter that controls
gravitational perturbation theory is $\kappa Q$. Just like the form
of soft-collinear QCD does not require the QCD coupling $g_s$ to be small,
since the power counting for soft and collinear fields is not
related to the size of $g_s$,
the construction of soft-collinear gravity applies in principle to
Planckian and trans-Planckian scattering energies.
We shall see below that higher-order terms
in the weak-coupling expansion, as well as higher-derivative interactions
that must be added to the Einstein-Hilbert action (\ref{eq:Einst})
to make it perturbatively renormalizable, are all suppressed for
collinear and soft gravitons, provided that $\lambda \kappa Q \ll 1$.
That is, we must require only that the transverse momentum scale
$\lambda Q$ is sufficiently below the Planck scale, but not the
scattering energies themselves.

The gauge is fixed by adding the term
\begin{equation}
\frac{b}{\kappa^{2}}\int\mathrm{d}^{4}x \left(  \partial_{\alpha}h_{\mu
}^{\alpha}-\frac{1}{2}\partial_{\mu}h\right)  \left(  \partial_{\beta}%
h^{\beta\mu}-\frac{1}{2}\partial^{\mu}h\right)
\end{equation}
to the action (\ref{eq:Einst}), where $b$ is an arbitrary, real, dimensionless
parameter. This corresponds to the covariant generalization of de~Donder
gauge, which is obtained for $b=1$. The graviton propagator
thus takes the form:
\begin{equation}
D_{\mu\nu,\alpha\beta}=\left\langle 0\left\vert T\,h_{\mu\nu}\left(  x\right)
h_{\alpha\beta}\left(  y\right)  \right\vert 0\right\rangle =i\kappa^{2}
\int\frac{\mathrm{d}^{4}p}{\left(  2\pi\right)  ^{4}}\frac{e^{-ip\cdot\left(
x-y\right)  }}{p^{2}+i0}\,\left(  P_{\mu\nu,\alpha\beta}+\frac{1-b}{b}%
\,S_{\mu\nu,\alpha\beta}\right)  , \label{eq:gravProp}%
\end{equation}
where
\begin{align}
P_{\mu\nu,\alpha\beta}  &  =\frac{1}{2}\left(  \eta_{\mu\alpha}\eta_{\nu\beta
}+\eta_{\mu\beta}\eta_{\nu\alpha}-\eta_{\mu\nu}\eta_{\alpha\beta}\right)
,\nonumber\\
S_{\mu\nu,\alpha\beta}  &  =\frac{1}{2 p^{2}}\left(  \eta_{\mu\alpha}p_{\nu
}p_{\beta}+\eta_{\mu\beta}p_{\nu}p_{\alpha}+p_{\mu}p_{\alpha}\eta_{\nu\beta
}+p_{\mu}p_{\beta}\eta_{\nu\alpha}\right)  .
\end{align}
Some components of the propagator (\ref{eq:gravProp}) vanish
identically,
\begin{equation}
D_{++,++}  =D_{++,+\perp}= D_{++,\perp\perp}=
D_{--,--}=D_{--,-\perp}=D_{--,\perp\perp}=0,
\label{eq:Dzero}%
\end{equation}
while the other independent components scale as follows:
\begin{align}
D_{++,+-}  &  \sim D_{+\perp,+\perp}\sim\lambda^{0},
\nonumber\\
 D_{++,-\perp}  & \sim D_{+\perp,\perp\perp}\sim
D_{+-,+\perp}\sim\lambda^{1},
\nonumber\\
D_{++,--}  &  \sim D_{+-,+-}\sim D_{+-,\perp\perp}\sim D_{+\perp,-\perp}\sim
D_{\perp\perp,\perp\perp}\sim\lambda^{2},
\label{eqDscale}\\
D_{+-,-\perp}  &  \sim D_{+\perp,--}\sim
D_{-\perp,\perp\perp}\sim\lambda^{3},
\nonumber\\
D_{+-,--} &\sim D_{-\perp,-\perp}\sim\lambda^{4}.\nonumber
\end{align}
Here the $\perp$ index denotes a contraction with the transverse
metric $\eta_{\mu\nu
}-(n_{+\mu}n_{-\nu}+n_{-\mu}n_{+\nu})/2$.

The scaling (\ref{eqDscale}) is only
consistent with the following counting rules for the components of the
field $h_{\mu\nu}$:
\begin{equation}
\begin{array}{lll}
h_{++}  \sim\lambda^{-1},\quad & h_{+\perp}\sim 1,\quad & h_{+-}\sim
\lambda, \\[0.2cm]
h_{--}  \sim\lambda^{3},\quad & h_{-\perp}\sim \lambda^2,\quad &
h_{\perp\perp}%
\sim\lambda.
\end{array}
\label{eq:Hscale}%
\end{equation}
Two points are to be made here. First, it is easy to see that any
combination $a h_{\mu\nu}+bh\,\eta_{\mu\nu}$ with $a\sim b\sim 1$ scales as
$h_{\mu\nu}$, since $h=h^\alpha_\alpha\sim \lambda$ and only
$\eta_{+-}$, $\eta_{-+}$, and $\eta_{\perp\perp}$ are non-zero.
This makes our consideration reparametrization
invariant; the gravitational field can be defined as the
linearized deviation of the contravariant metric density $\sqrt{-g}g^{\mu\nu}$
or the vierbein field $e_{\mu}^{\left(  a\right)  }$ and so on. Second,
the power counting presented above depends only on the
number of \textquotedblleft$-$\textquotedblright\ components $N_{-}$ and
\textquotedblleft$\perp$\textquotedblright\ components $N_{\perp}$.
The scaling of the propagator components $D_{\mu\nu,\alpha\beta}$
takes the form
$D_{\mu\nu,\alpha\beta}\sim\lambda^{2N_{-}+N_{\perp}-2}$,
and
\begin{equation}
h_{\mu\nu}\sim\lambda^{2N_{-}+N_{\perp}-1}
\end{equation}
for the gravitational field.
Hence the \textquotedblleft$\perp$\textquotedblright\ index contributes as
$\lambda$, the \textquotedblleft$-$\textquotedblright\ index
as $\lambda^{2}$, the \textquotedblleft$+$\textquotedblright\  index as
$\lambda^{0}$, and there is one additional factor $\lambda^{-1}$ for
every $h_{\mu\nu}$. An immediate and somewhat unusual consequence of
this is that the collinear metric component $h_{++}$ is
{\em enhanced} in the small power-counting parameter $\lambda$.
This is the first important difference between collinear gravitational and
gauge fields.

It is also easily checked that the contraction of $h_{\mu\nu}$ with any
four-vector $V_\nu$, collinear to the same $n_{-}$, that is,
with scaling (\ref{eq:ColScale}),
yields an additional power suppression:%
\begin{equation}
h^{\mu\nu}V_{\nu}=\frac{1}{2}\left(  h_{+}^{\mu}V_{-}+h_{-}^{\mu}%
V_{+}\right)  +h_{\perp}^{\mu}V_{\perp}\sim \lambda\, V^{\mu},
\label{eq:Acontract}
\end{equation}
which holds for every component $\mu$ separately. This provides the
second main difference between the collinear metric field $h_{\mu\nu}$ and
the collinear gauge field $A_{\mu}$. The coupling to matter is given
through an action $S_{m}$ by
\begin{equation}
S_{int}=\int\mathrm{d}^{4}x\,\left(  A^{\mu}\frac{\delta S_{m}}{\delta A^{\mu
}}+h_{\mu\nu}\,\frac{\delta S_{m}}{\delta h_{\mu\nu}}\right)  .
\end{equation}
For example, for the coupling to a fermion,
$\delta S_{m}/\delta A^{\mu}\propto j_\mu = \bar\psi\gamma_\mu\psi$,
and $\delta S_{m}/\delta h_{\mu\nu} \propto T^{\mu\nu} =
\bar\psi\gamma^\mu i\partial^\nu \psi $.
In comparison to $\delta S_{m}/\delta A^{\mu}$, the variation $\delta
S_{m}/\delta h_{\mu\nu}$ has an additional Lorentz index. If the matter is
also a collinear field then the additional Lorentz index implies an additional
contraction of the type (\ref{eq:Acontract}). This yields
power suppression in $\lambda$ relative to the coupling to the collinear
gauge field, exactly in line with our qualitative discussion. A
consequence of this is that the $\lambda$-expansion of the
collinear matter-coupling to gravitation almost coincides with the
weak-field expansion. This will be illustrated in more detail for the
expansion of the fermion Lagrangian in the next section.

Soft modes of the gravitational field can be also estimated in the
same manner. The metric is decomposed into collinear $h_{\mu\nu}$ and soft
$s_{\mu\nu}$ fields according to
\begin{equation}
g_{\mu\nu}=\eta_{\mu\nu}+h_{\mu\nu}+s_{\mu\nu}.
\end{equation}
Derivatives acting on the soft field counted as the corresponding soft
momentum (\ref{eq:SoftScale}). The components (\ref{eq:Dzero}) vanish also
for the soft field, while all other projections scale as $\lambda^{4}$.
Therefore, any component of $s_{\mu\nu}$ scales as $\lambda^{2}$.

\subsection*{3 Soft-collinear Lagrangian}

As an example, we consider the massless spinor field coupled to the
gravitational field through the action
\begin{equation}
S_{m}=\frac{1}{2} \int\mathrm{d}^{4}x\,\sqrt{-g}~\left[
\bar{\psi} \,E_{(a)}^{\mu} \gamma^{a}
\,\left(i\overrightarrow{D}_{\mu}\psi\right)  -\left(
\bar{\psi}\,i\overleftarrow{D}_{\mu}\right)
E_{(a)}^{\mu} \gamma^{a}
\psi\right].
\label{eq:spinS}%
\end{equation}
The covariant derivatives act on fermions as
\begin{equation}
\left(  \overrightarrow{D}_{\mu}\psi\right)  =\partial_{\mu}\psi
-\frac{i}{2}\Sigma^{ab}\gamma_{abc}e_{\mu}^{(c)}\psi,\quad\left(  \bar
{\psi}\overleftarrow{D}_{\mu}\right)  =~\partial_{\mu}\bar{\psi}%
+\frac{i}{2}\bar{\psi}~\Sigma^{ab}\gamma_{abc}~e_{\mu}^{(c)},
\end{equation}
where $\Sigma^{ab}=\frac{i}{4}\left[  \gamma^{a},\gamma^{b}\right]$.
The vierbein field $e_{\mu}^{\left(  a\right)  }$, its
inverse $E_{\left(  a\right)  }^{\mu}$ and the spin connection
$\gamma_{abc}\ $ are defined as
\begin{equation}
\eta_{ab}\,e_{\mu}^{\left(  a\right)  }e_{\nu}^{\left(  b\right)  }=g_{\mu\nu
},\quad E_{\left(  a\right)  }^{\mu}e_{\nu}^{\left(  a\right)  }=\delta_{\nu
}^{\mu},\quad\gamma_{abc}=E_{\left(
b\right)  }^{\mu}E_{\left(  c\right)  }^{\nu}
D_\nu e_{\left(  a\right)  }{}_{\mu},
\end{equation}
respectively.
The weak field expansion of the action (\ref{eq:spinS}) results in the
Lagrangian
\begin{equation}
{\cal L}_m =
\left(\delta_{\nu}^{\mu}-H_{\nu}^{\mu}\right)
\,\bar{\psi}\,\gamma^{\nu} i\!\overleftrightarrow{\partial}_{\!\!\mu}\psi,
\qquad\overleftrightarrow{\partial}_{\!\!\mu} =
\frac{1}{2}\left(\overrightarrow{\partial}_{\!\!\mu}-\overleftarrow{\partial
}_{\!\!\mu}\right)  ,
\label{eq:spinL}
\end{equation}
where
\begin{equation}
H_{\nu}^{\mu}=\frac{1}{2}\left(  h_{\nu}^{\mu}-h\,\delta_{\nu}^{\mu}\right)  .
\end{equation}

Integrating out the $\zeta$ component of the spinor field (see the definition
(\ref{eq:PsiScale})) and using the power-counting rules (\ref{eq:PsiScale})
and (\ref{eq:Hscale}), we expand the matter Lagrangian in
powers of $\lambda$, ${\cal L}={\cal L}^{(0)}+{\cal L}^{(1)}+\ldots$,
where the superscript means that ${\cal L}^{(n)}\sim\lambda^n {\cal L}^{(0)}$.
The leading term and the first power correction in the purely
collinear Lagrangian are found to be
\begin{align}
{\cal L}^{(0)}
& =\bar{\xi}\,\frac{\not n  _{+}}{2}\left(  i\overrightarrow{\partial
}_{\!-}+i\overrightarrow{\partial}_{\!\perp}
\frac{1}{i\overrightarrow{\partial}_{\!+}}\,
i\overrightarrow{\partial}_{\!\perp}\right) \xi,
\label{eq:L0expand}\\[0.2cm]
{\cal L}^{(1)}  &  =
-H_{-}^{\mu}\,\,\bar{\xi}\,\frac{\not n  _{+}}%
{2}\left(  i\overleftrightarrow{\partial}_{\!\!\mu}\right)  \xi
+i\,\bar{\xi}\,
\frac{\not n  _{+}}{2}\left[-\overleftarrow{\not \partial }_{\!\perp}\frac
{1}{\overrightarrow{\partial}_{\!+}}\left(  H_{+}^{\mu}\overrightarrow{\partial
}_{\!\mu}+\frac{1}{2}H_{+,\mu}^{\mu}\right)  \frac{1}{\overrightarrow{\partial
}_{\!+}}\overrightarrow{\not \partial }_{\!\perp}\right.
\nonumber\\
&  \left.  +\overleftarrow{\not \partial }_{\!\perp}\frac{1}{\overrightarrow
{\partial}_{\!+}}\left(  H_{\nu}^{\mu}\overrightarrow{\partial}_{\!\mu
}+\frac{1}{2}H_{\nu,\mu}^{\mu}\right)  \gamma_{\perp}^{\nu}%
+\gamma_{\perp}^{\nu}\left(  \overleftarrow{\partial}_{\!\mu}H_{\nu
}^{\mu}+\frac{1}{2}H_{\nu,\mu}^{\mu}\right)  \frac{1}{\overrightarrow
{\partial}_{\!+}}\overrightarrow{\not \partial }_{\!\perp}\right]  \xi.
\label{eq:L1expand}
\end{align}
The inverse derivative $\left(  \overrightarrow{\partial}_{\!+}\right)
^{-1}$ should be read as \cite{Beneke2002}
\begin{equation}
\frac{1}{i\partial_{+}+i0}\,\phi(x)=
-i\int_{-\infty}^{0}\mathrm{d}u\,\phi\left(x+u n_{+}\right),
\end{equation}
and an index after the comma denotes an ordinary partial derivative.
The leading-power Lagrangian ${\cal L}^{(0)}$ has a similar
form as in \cite{Beneke2002} as far as derivatives are concerned, but
contrary to the gauge-field case it is non-interacting. All purely
collinear interactions are power-suppressed!
The fact that the interaction of collinear fields with collinear gravitons is
power suppressed is readily apparent from the weak-field
expansion~(\ref{eq:spinL}). Indeed, the ``current''
$\bar{\psi}\,\gamma^{\nu} i\!\overleftrightarrow{\partial}_{\!\!\mu}\psi$
carries an additional collinear four-vector index, which is
contracted with the metric field.
As was stressed above (see (\ref{eq:Acontract})),
this yields power suppression.

A similar reasoning shows that the ``decoupling'' of collinear gravitons
in the leading-power Lagrangian is independent of the type
of field, since any collinear Lagrangian contains only collinear
four-vectors, which must be contracted at least once with the metric
field. This applies, in particular, to the self-interaction of collinear
gravitons. Likewise, higher-order corrections to the weak-coupling
expansion are further and further suppressed in the purely
collinear sector, since additional powers of $h_{\mu\nu}$ require
additional contractions, while $h=h^\mu_\mu$ is itself of order
$\lambda$. The same suppression applies to higher-derivative
terms that should be added to the Einstein-Hilbert action
(\ref{eq:Einst}) to make it perturbatively renormalizable. Soft
derivatives are power-suppressed by construction, while the
collinear derivative terms always come with an additional collinear
contraction. This shows that the applicability of the
effective theory of soft-collinear quantum gravity is restricted
to transverse momenta rather than energies
smaller than the Planck scale.

In contrast to the purely collinear interaction, the soft-collinear one is not
power suppressed. Writing $S_{\nu}^{\mu}=\frac{1}{2}
\left(  s_{\nu}^{\mu}-s\,\delta_{\nu}^{\mu}\right)$, the leading
soft interaction in (\ref{eq:spinL}) is contained in
\begin{equation}
\left(  \delta_{\nu}^{\mu}-S_{\nu}^{\mu}\right)  \,\bar{\xi}\,
\frac{\not n  _{+}}{2}n_{-}^{\nu}
\,i\!\overleftrightarrow{\partial}_{\!\!\mu}\xi=\underset
{\sim\lambda^{4}}{\underbrace{j_{-}}}-\frac{1}{2}\,\underset{\sim\lambda^{4}%
}{\underbrace{S_{--}\,j_{+}}}-\underset{\sim\lambda^{5}}{\underbrace
{S_{-\perp}\,\,j_{\perp}}}-\frac{1}{2}\,\underset{\sim\lambda^{6}}%
{\underbrace{S_{-+}\,j_{-}}}, \label{eq:localSoft}%
\end{equation}
where
\begin{equation}
j_{\mu}=\bar{\xi}\,\frac{\not n  _{+}}{2}\,i\!
\overleftrightarrow{\partial}_{\!\!\mu}\,\xi  .
\end{equation}
In the expression (\ref{eq:localSoft}), we indicate the $\lambda$-scaling
with underbraces, which identifies the second term on the right-hand side
as the only leading-power interaction. The structure of this interaction
is very similar to the corresponding gauge field interaction
$\bar{\xi}\,\frac{\not n  _{+}}{2} \xi A_{s-}$. In further analogy with
the soft-collinear gauge interaction, soft fields interacting with
collinear fields must be multipole-expanded in their position argument
to achieve a homogeneous $\lambda$-expansion~\cite{Beneke2002}. If
the $\lambda$-expansion is limited to the leading order, it is sufficient to
replace the argument of the soft field $x^{\mu}$ by $n_{-}^{\mu}x_{+}/2$,
where $n_{-}^{\mu}$ is the reference direction.\footnote{Note the change
of notation: the four-vector
$n_{-}^{\mu}x_{+}/2$ was called $x_-^\mu$ in~Ref.~\cite{Beneke2002},
while here we use $x_+^\mu$, or simply $x_+$.}  Therefore, the leading-order
Lagrangian of the collinear fermion field interacting with the soft
gravitational field takes the form:
\begin{equation}
{\cal L}_{c+s}^{(0)}=\bar{\xi}\,\frac{\not n  _{+}}{2}\left[
i\!\overleftrightarrow{\partial}_{\!\!-}-\frac{1}{4}s_{--}\left(  x_{+}\right)
\,i\!\overleftrightarrow{\partial}_{\!\!+}
+i\overrightarrow{\partial}_{\!\perp}
\frac{1}{i\overrightarrow{\partial}_{\!+}}i\overrightarrow{\partial}_{\!\perp
}\right]  \xi.
\label{eq:softLag}
\end{equation}
We emphasize again that the only component remaining is $s_{--}$, similar
to the leading SCET Lagrangian which contains only the $A_{-}$ soft
component. In both situations this corresponds to the eikonal approximation
for soft fields. We also note that on functions $f(x_+)$ of
$x_+ = n_+ x$ only the minus-component $\partial_-$ of $\partial_\mu$
is non-vanishing, hence $\partial_+$ in (\ref{eq:softLag}) does not
operate on the metric field.

\subsection*{4 Collinear factorization}

\begin{figure}[ptb]
\noindent
\parbox[c]{.52\textwidth}{\centering
\includegraphics[width=.3\textwidth]{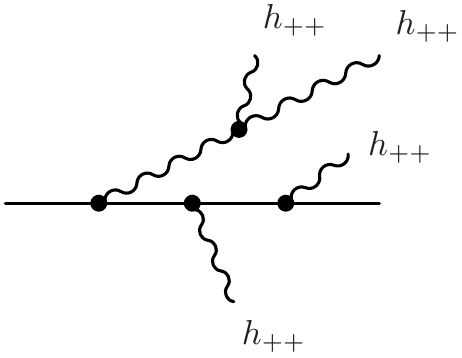}\par }
\parbox[c]{.46\textwidth}{\centering
\includegraphics[width=.25\textwidth]{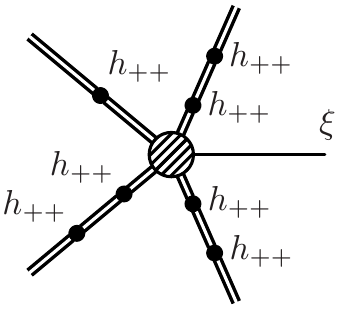}\par }
\vskip0.5cm
\parbox[t]{\textwidth}{
\caption{a) (left panel) Collinear splittings with $h_{++}$-ends
generated by three-point
vertices. b) (right panel) Source that generates non-collinear lines
(double lines) to which the enhanced $h_{++}(x_{-})$ field can couple.}
\label{fig:pic1}}
\end{figure}

{}From the fore-going it might be concluded that collinear factorization
is trivial, since collinear interactions are power-suppressed, and
that nothing remains to be done except for checking the collinear
limit of the leading-power soft (eikonal) interaction as was done in
Ref.~\cite{Weinberg1965}. This is not quite true. As discussed in
Ref.~\cite{Akhoury2011}, although the purely collinear sector is power
suppressed, its interaction with a non-collinear environment is not.
In the present framework, this follows from the negative $\lambda$-scaling
of the  $h_{++}$ metric component, which implies that its coupling
to an energetic (or massive) particle not collinear with $h_{\mu\nu}$
should be counted as $\lambda^{-1}$. This negative scaling can compensate the
$\lambda$-suppression of the interaction of collinear matter $\xi$
with collinear gravitons, and of collinear graviton self-interactions,
thus generating diagrams with a leading-power collinear singularity.
The situation is illustrated in Fig.~\ref{fig:pic1}. The left panel
shows an example of a collinear splitting diagram. The $\lambda$-suppression
of the collinear vertices can be compensated only if every line
ends with the $h_{++}$ field. In order not to incur further
suppression these ends must be tied to non-collinear (with respect to
the direction of $h_{++}$) lines, as shown by the double lines
in the right panel. The $\xi$ line in this figure stands for the
entire collinear sector on the left, which, together with the
non-collinear lines, is generated by an unspecified source
operator, represented by the shaded blob.\footnote{The
right panel in Fig.~\ref{fig:pic1} corresponds to
what is referred to as the ``rest'' of the diagram in
Ref.~\cite{Akhoury2011}.}

It is instructive to compare this situation to SCET for gauge
interactions. The $A_+$ component of the gauge field scales
as $\lambda^0$, but it is not enhanced by a negative scaling
dimension. For SCET to be a useful effective theory,
the interactions of the unsuppressed $A_+$ field
must be controlled to all orders in the $A_+$ field, separately in
every order in the $\lambda$-expansion. This is accomplished
by showing that $A_+$ appears in SCET only through collinear
Wilson lines $W_c$~\cite{Bauer2002,Beneke2002}.
Clearly, in the case of gravity the coupling of the $h_{++}$ field
must be also controlled to all orders. In contrast to SCET,
the collinear Lagrangian to all orders in the $\lambda$-expansion
is now potentially relevant, since the $\lambda$-suppression of collinear
interactions can be compensated by the negative $\lambda$-scaling
of the $h_{++}$ field coupling to the sources.
Hence, we need to factorize $h_{++}$ from
the sources without making use of the
collinear Lagrangian to any finite order. Below we
show how this can be done in complete generality.

The difficulty with negative scaling of $h_{++}$ can be avoided by
adopting the ghost-free light-cone
gauge~\cite{Scherk1975, Kaku1975, Leibbrandt1987}. The propagator
in this gauge reads
\begin{align}
\left\langle 0\left\vert T\,h_{\mu\nu}\left(x\right) h_{\alpha\beta}\left(
y\right)  \right\vert 0\right\rangle
&  =i\kappa^{2}\int\frac{\mathrm{d}^{4}p}{\left(2\pi\right)^{4}}
\frac{e^{-ip\cdot\left(x-y\right)}}{p^{2}+i0}
\frac{1}{2}\left(  \rho_{\mu\alpha}\rho_{\nu\beta}+\rho_{\mu\beta
}\rho_{\nu\alpha}-\rho_{\alpha\beta}\rho_{\mu\nu}\right),
\nonumber \\
\rho_{\mu\nu}
&  =\eta_{\mu\nu}-\frac{p_{\mu}n_{+\nu}+p_{\nu}n_{+\mu}}{p\cdot
n_{+}}.
\label{eq:LCgauge}
\end{align}
The components $h_{+\mu}$ are now non-propagating.
Thus we can exclude the unphysical components $h_{+\mu}$ from the
path-integral so that only metric components with positive
$\lambda$-scaling remain. The absence of collinear singularities
in gauge-invariant observables is therefore guaranteed.
However, to gain a better understanding of factorization,
we use here another way to exclude the interaction with $h_{++}$,
which does not refer to a non-covariant gauge.
Note that in SCET the unsuppressed gauge field component
$A_+$ can likewise be eliminated in light-cone gauge.
But contrary to gravity this does not eliminate collinear divergences,
since the collinear Lagrangian contains leading-power
interactions with the other gauge field components.

Let us therefore consider the zero-momentum insertion of a
generally covariant, local operator
\begin{equation}
O\left(  \phi_{1},..,\phi_{n},\xi,g\right)  =\int\mathrm{d}^{4}x\,\sqrt
{-g}\,P\left(  \phi_{1},..,\phi_{n},\xi,g\right)  , \label{eq:O}%
\end{equation}
where $P$ is the ``source'' of any number of fields including the
collinear one, $\xi$, that is, a polynomial containing fields $\phi_i$,
$\xi$, covariant derivatives, vierbeins, and the metric tensor at the
same space-time point.
The operator (\ref{eq:O}) corresponds to the vertex
depicted as shaded blob in Fig.~\ref{fig:pic1}. Each field
$\phi_{i}$ in (\ref{eq:O}) also interacts with $h_{++}$ through
some Lagrangian ${\cal L}_{i}(\phi_{i},g)$.

An interaction vertex of $h_{++}$ with a non-collinear field $\phi_i$
should be classified as ``hard'', since the
momentum transfer is of the order of hard scale $Q$.
Large virtualities $Q^2$ can be achieved only through the
large-momentum component $p_+$ of the $h_{++}$ line at the vertex.
In position space it therefore suffices to consider the interaction with
$h_{++}\left(x_{-}\right)$, where $x_-^\mu$ is defined as
$n_{+}^{\mu}x_{-}/2$, thus
neglecting the dependence of $h_{++}(x)$ on $x_{+}$ and $x_{\perp}$.
The intermediate, highly virtual non-collinear lines
in Fig.~\ref{fig:pic1} should be contracted and represented
as an effective vertex in soft-collinear gravity.
We shall demonstrate the cancellation of collinear singularities
in the sum of diagrams generated by
these effective vertices in two steps: first, we show that
through a field redefinition
the interactions with $h_{++}\left(  x_{-}\right)$ can be collected
into universal factors in $O$ and ${\cal L}_{i}$ that take the form
of a gauge (coordinate) transformation. This should be expected for an
unphysical field component, and, as we will see shortly, this
transformation is particularly simple when the metric
field is $h_{++}\left(  x_{-}\right)$.
This step is very similar to the factorization
of collinear modes in QCD and the field redefinition exploited in
Refs.~\cite{Bauer2002,Beneke2003}.
However, in contrast to QCD, in the second step we demonstrate that
these factors cancel out due to translation invariance.

Since in the following we are concerned only with the $h_{++}$ metric
component, we may specialize the
space-time (\ref{eq:weakG}) to
\begin{equation}
g_{\mu\nu}=\eta_{\mu\nu}+\frac{n_{-\mu}n_{-\nu}}{4}h_{++}\left(x_{-}\right),
\label{eq:LCmetric}
\end{equation}
which can also be denoted as follows:
\begin{equation}
\hat{g}=\hat{\eta}+\hat{h},\quad\text{so that}\quad\hat{h}^{2}=0,\quad
\operatorname*{tr}\hat{h}=0.
\end{equation}
The algebra of the matrix  $\hat{h}$
resembles the algebra of a single Grassmann number. Any
analytic function of $\hat{h}$ is a linear function, hence we can
find the exact expressions for the contravariant metric tensor and
the metric determinant:
\begin{equation}
\hat{g}^{-1}=\hat{\eta}-\hat{h},\qquad\det\hat{g}=-1.
\end{equation}
It is straightforward to find the vierbeins
\begin{equation}
e_{\beta}^{\left(  \alpha\right)  } =\delta_{\beta
}^{\alpha}+\frac{n_{-\beta}n_{-}^{\alpha}}{8}\,h_{++}\left(  x_{-}\right)
,
\qquad
E_{\left(  \alpha\right)  }^{\beta}
=\delta_{\alpha}^{\beta}-\frac{n_{-\alpha}n_{-}^{\beta}}{8}%
\,h_{++}\left(  x_{-}\right),
\label{eq:vierbeinspecial}
\end{equation}
and the affine connection
\begin{equation}
\Gamma^\lambda_{\mu\nu} = \frac{1}{16}
n_-^\lambda n_{-\mu} n_{-\nu} \,\partial_+h_{++}(x_-).
\label{eq:affinespecial}
\end{equation}
One further verifies that the covariant derivative of
the vierbein is zero
\begin{equation}
D_\nu e^{(a)}_\mu = \partial_\nu e^{(a)}_\mu -
\Gamma^\lambda_{\nu\mu} e^{(a)}_\lambda =0,
\label{eq:Dezero}
\end{equation}
which implies that the spin connection $\gamma_{abc}$ vanishes.
{}From (\ref{eq:affinespecial}) it also follows that the
Riemann tensor $R^\mu_{\phantom{\mu}\nu\rho\sigma}$ vanishes
so the space-time (\ref{eq:LCmetric}) is flat. The
transformation to the global inertial frame with
coordinates $y^\mu$ is the local translation
\begin{equation}
y^{\mu}=x^{\mu}+\frac{n_{-}^{\mu}}{2}\,w\left(  x_{-}\right)  ,\qquad
w\left(  x_{-}\right)  =\frac{1}{4}\int_{-\infty}^{x_{-}}\mathrm{d}%
x_{-}^{\prime}h_{++}\left(  x_{-}^{\prime}\right)  ,
\end{equation}
so that the metric (\ref{eq:LCmetric}) can be obtained via the standard
relation
\begin{equation}
g_{\mu\nu}\left(  x\right)  =
\eta_{\alpha\beta} \,\frac{\partial y^{\alpha}}{\partial x^{\mu}}
\frac{\partial y^{\beta}}{\partial x^{\nu}}.
\end{equation}

For the first step in the demonstration of the cancellation of
collinear singularities, we consider a general covariant Lagrangian
${\cal L}_{i}(\phi_{i},g, D)$ for some matter field $\phi_{i}$ in the
space-time (\ref{eq:LCmetric}). Due to general covariance, we
could go to the ``flat'' coordinates $y$. Then
$g\to \eta$ and $D\to\partial$ in ${\cal L}_i$, and
all interactions with $h_{++}$ disappear, which already proves
the decoupling in the Lagrangian.

It is, however, instructive to show the decoupling in the original
coordinate system representing all transformations as field redefinitions.
Let $\hat \Gamma$ be the $4\times 4$ matrix with entries
$\Gamma^\mu_{+\nu}$, and define the parallel transport
matrix $\hat U$ through
\begin{equation}
U^{\mu}_{\nu}\left(x_{-}\right)  =\left[  \mathrm{P}\exp\left(
\int_{-\infty}^{x_{-}}\mathrm{d}x_{-}^\prime\,
\frac{1}{2}\,\hat \Gamma(x_-^\prime)\right)\right]_{\nu}^{\mu},
\end{equation}
and the collinear Wilson-line operator as follows:
\begin{eqnarray}
&& W\left(  x_{-}\right)  =\exp\left[  -\frac{i}{4}\int_{-\infty}^{x_{-}%
}\mathrm{d}x_{-}^{\prime}h_{++}\left(  x_{-}^{\prime}\right)  \frac{i}%
{2}\partial_{-}\right]
= \exp\left[ w(x_-)\frac{1}{2}\,\partial_-\right]   ,\nonumber\\[0.2cm]
&& W^{-1}\left(  x_{-}\right)  =\exp\left[
\frac{i}{4}\int_{-\infty}^{x_{-}}\mathrm{d}x_{-}^{\prime}h_{++}\left(
x_{-}^{\prime}\right)  \frac{i}{2}\partial_{-}\right].
\label{eq:W}%
\end{eqnarray}
The field is assumed to vanish at $-\infty$, so that the integrals are
convergent. Note that $\partial_-$ does not operate on $h_{++}(x_-)$,
since it acts only on $x_+= n_+ x$. Also
\begin{equation}
\frac{1}{2}\,\partial_- =
\frac{n_+^\mu}{2}\, \partial_\mu =
\frac{\partial}{\partial(n_- \cdot x)}
= \frac{\partial}{\partial x_+}.
\end{equation}
Due to the nilpotency of
$\hat \Gamma$, which follows from (\ref{eq:affinespecial}),
the exponential is in fact linear, and the relations
\begin{equation}
U_{\mu}^{\alpha} = e_{\mu}^{\left(  \alpha\right)  },
\qquad
\left(  U^{-1}\right)  _{\alpha
}^{\mu}=E_{\left(  \alpha\right)  }^{\mu}
\label{eq:U=E}
\end{equation}
hold. At the operator level, the Wilson lines acts as
\begin{equation}
W\phi\left(  x\right)  W^{-1} =
\phi\left(  x\right) +w\left(x_{-}\right)
\left[  \frac{\partial}{\partial x_{+}},\phi\left(  x\right)
\right]
+\frac{1}{2!}w^{2}\left(  x_{-}\right)  \left[  \frac{\partial
}{\partial x_{+}},\left[  \frac{\partial}{\partial x_{+}},\phi\left(
x\right)  \right]  \right]  +\ldots
\label{eq:Hadamard}
\end{equation}
The Wilson line (\ref{eq:W}) has the form of the translation
operator to the flat coordinate point,
\begin{equation}
W\phi\left(  x\right)  W^{-1}
= \sum_{n=0}^\infty \frac{1}{n!}\,w(x_-)^n\,\frac{\partial^n}
{\partial x_+^n}\,\phi(x)
= \phi\big(x+\frac{n_-}{2} w(x_-)\big) = \phi(y).
\label{eq:Hadamard2}
\end{equation}

This motivates the following field redefinition. Let
$\phi_i^{\nu_{1}\nu_{2}..\nu_{n}}$ be an arbitrary field
with $n$ generally covariant indices. Then define
\begin{equation}
\phi_{i}^{\prime\,\mu_{1}\mu_{2}..\mu_{n}}=W^{-1}U_{\nu_{1}}^{\mu_{1}}U_{\nu
_{2}}^{\mu_{2}}..U_{\nu_{n}}^{\mu_{n}}\phi_i^{\nu_{1}\nu_{2}..\nu_{n}}W.
\label{eq:Fredef}
\end{equation}
The interpretation of this expression is clear: due to (\ref{eq:U=E})
$U_{\nu_{1}}^{\mu_{1}}U_{\nu_{2}}^{\mu_{2}}..U_{\nu_{n}}^{\mu_{n}}
\phi_i^{\nu_{1}\nu_{2}..\nu_{n}}$ is the field in the
local inertial frame, which in the case at hand is global.
The $W$ operators perform the translation from $x$ to $y$,
hence the redefined
field should correspond to the decoupled field in $y$ coordinates.
Indeed, we now show that in terms of the redefined fields,
\begin{equation}
{\cal L}_i(\phi_i,g,D) = W {\cal L}_i(\phi_i^\prime,\eta,\partial) \,W^{-1},
\label{eq:Lidecoup}
\end{equation}
which implies
\begin{equation}
\int\mathrm{d}^{4}x\,\underbrace{\sqrt{-g}}_{=1}
\,{\cal L}_i\left(\phi_i,g,D\right)
=\int\mathrm{d}^{4}x\,W {\cal L}_i\left(
\phi_i^{\prime},\eta,\partial\right)  W^{-1}
=\int\mathrm{d}^{4}x\,{\cal L}_i\left(
\phi^{\prime}_i,\eta,\partial\right),
\label{eq:Lprop}
\end{equation}
where, in the last equality, the disappearance of the factors $W$
and $W^{-1}$ is a consequence of translation invariance, i.e.,
energy-momentum conservation, or simply of dropping total derivative
terms. This expresses the decoupling
of $h_{++}$ from the non-collinear lines in the original $x$-coordinates,
since the metric field no longer appears in the action.

To prove (\ref{eq:Lidecoup}) we express the field
$\phi_i^{\nu_{1}\nu_{2}..\nu_{n}}$ in terms of the primed field. We
note that $W$ commutes with $U$, since $U$ depends only on $x_-$. It is
straightforward to show the identities
\begin{eqnarray}
&& \partial_\nu \,[W\ldots ] = U^\rho_\nu \,W\partial_\rho [\ldots],
\label{eq:idW1}\\[0.2cm]
&&
D_\nu \,\big[(U^{-1})_{\mu_1}^{\nu_1}\ldots
(U^{-1})_{\mu_n}^{\nu_n} \ldots \big]
=  (U^{-1})_{\mu_1}^{\nu_1}\ldots
(U^{-1})_{\mu_n}^{\nu_n} \,\partial_\nu \big[\ldots\big],
\label{eq:idW2}
\end{eqnarray}
where the second follows from (\ref{eq:Dezero}) or the vanishing of the
spin connection. With the help of these identities we convert all
covariant derivatives acting on $\phi_i^{\nu_1\nu_2..\nu_n}$
into ordinary ones. Furthermore, after
applying (\ref{eq:idW1}) the $W$ operators
and their inverses arising from a product of $\phi_i$ fields
can be cancelled except for one $W$ on the left and one $W^{-1}$
on the right. Since all generally covariant indices in the original
Lagrangian must be contracted by $g_{\mu\nu}$ (fields) and
$g^{\mu\nu}$ (covariant derivatives), the $U$ and $U^{-1}$ factors
must necessarily be multiplied in the form
\begin{equation}
g_{\mu\nu}\,(U^{-1})^{\mu}_{\alpha}(U^{-1})^{\nu}_{\beta} =
\eta_{\alpha\beta},
\qquad
g^{\mu\nu}\,U_{\mu}^{\alpha}U_{\nu}^{\beta} =
\eta^{\alpha\beta},
\end{equation}
where the equalities follow from the identity (\ref{eq:U=E})
of $U$ with the vierbein. This removes any appearance of the
metric and $U$ from the Lagrangian, thus
completing the proof of (\ref{eq:Lidecoup}).

At this point after the field redefinition, the
interaction with $h_{++}\left(x_{-}\right)$ remains only in the source
operator (\ref{eq:O}). However, nothing above was special to
the Lagrangian interactions, and we can apply the same field
redefinition to the collinear fields $\xi$ and $h$.
Hence, the relations (\ref{eq:Lidecoup}), (\ref{eq:Lprop}) are also
valid for the operator (\ref{eq:O}), that is
\begin{equation}
P\left(  \phi_{1},..,\phi_{n},\xi,g,D\right)  =
WP\left(\phi_{1}^{\prime},..,\phi_{n}^\prime,\xi^\prime,\eta,\partial \right)
W^{-1},
\label{eq:P}
\end{equation}
and $O\left(  \phi_{1},..,\phi_{n},\xi,g\right) =
O\left(\phi_{1}^{\prime},..,\phi_{n}^\prime,\xi^\prime,\eta\right)$
due to translation invariance. This shows the decoupling of
the dangerous $h_{++}$ field with negative scaling dimension
from the sources and the non-collinear fields. Due to the power-suppression
of collinear self-interactions, this excludes the presence of
collinear divergences in physical processes.

We briefly compare the above result to
Refs.~\cite{Weinberg1965,Akhoury2011}. As mentioned above,
Weinberg \cite{Weinberg1965} works in the eikonal approximation
and shows diagrammatically that no additional collinear singularities
arise when any of the non-infrared particles (in our terminology,
lines emanating from the source in Fig.~\ref{fig:pic1}) becomes massless,
provided momentum is conserved at the source vertex,
which in our treatment corresponds to the use of
translation invariance and a zero-momentum source.

The proof of collinear cancellations in the present paper
is not restricted to the eikonal approximation. The general
case was also analyzed recently by Akhoury et al.~\cite{Akhoury2011}
employing diagrammatic factorization methods. Their conclusion
is equivalent to ours, but their proof contains the additional
result that only collinear three-point but no higher-point
vertices in the branchings depicted in the left Figure~\ref{fig:pic1}
can lead to diagrams with collinear divergences. No such
statement follows from the present treatment. This stronger statement
holds only in de~Donder gauge ($b=1$) as was assumed in
Ref.~\cite{Akhoury2011}, and not in the general covariant
gauge (\ref{eq:gravProp}). In de~Donder gauge, any non-zero
component of the momentum-space graviton propagator contributes as
$1/p^{2}\sim\lambda^{-2}$ to the collinear degree of divergence of
a given diagram. In the general covariant gauge the
propagator (\ref{eq:gravProp}) contains more singular terms such as
$p_{+}p_{+}/\left(p^{2}\right)^2\sim\lambda^{-4}$, and the power-counting
formula must be modified. In de~Donder gauge, the structure of
the collinear ``tree'' in  Figure~\ref{fig:pic1} is rather special.
Since the branches must end in $h_{++}$, and since in
de~Donder gauge (contrary to the general covariant gauge,
see (\ref{eqDscale})) the only non-vanishing component of
$D_{++,\alpha\beta}$ is $D_{++,--}$, the relevant component of
a triple vertex to which two ending branches attach is
$h_{--}h_{--}h_{++} p_{+} q_{+}$, where $p$, $q$
are two collinear momenta at the vertex. Hence, the
internal propagator at this vertex has ++ indices. We can now
repeat this argument to conclude that the ++ index is transported
through the tree. It is now clear why four- and
higher-point vertices cannot contribute, since this would
lead to at least six minus indices at the vertex which cannot
all be paired with plus indices.
In general covariant gauge this argument fails,
since the propagator (\ref{eqDscale}) can link
$h_{++}$ with $h_{--}$ as well as with $h_{+-}$ or $h_{-\perp}$.
Therefore, the class of vertices involved in collinear ``trees''
is larger in general. The scaling
(\ref{eq:Hscale}) implies that an $n$-graviton vertex containing two
collinear momenta scales as $\lambda^{n-2}$. However, this suppression is
compensated by the corresponding growth of the number of $h_{++}$
lines attached to the non-collinear part of a diagram. Therefore,
in the general covariant gauge
the ``trees'' can contain $n$-graviton vertices with $n>3$.

\subsection*{5 Soft factorization}

A similar reasoning shows the factorization of soft
graviton interactions, but in this case there is no cancellation.
The soft Wilson lines can be defined
in a similar manner as (\ref{eq:W}), but
since the leading-power soft-collinear
Lagrangian  (\ref{eq:softLag}) contains only the
soft graviton field $s_{--}(x_+)$, the appropriate expression
reads
\begin{align}
& Z_{n}\left(  x_{+}\right) =\exp\left[  -\frac{i}{4}\int_{-\infty}^{x_{+}%
}\mathrm{d}x_{+}^{\prime}\,s_{--}\left(  x_{+}^{\prime}\right)  \frac{i}%
{2}\partial_{+}\right] = \exp\left[z(x_+) \frac{1}{2}\partial_+
\right],\nonumber\\[0.12cm]
& Z_{n}^{-1}\left(  x_{+}\right)  =\exp\left[
\frac{i}{4}\int_{-\infty}^{x_{+}}\mathrm{d}x_{+}^{\prime}\,s_{--}\left(
x_{+}^{\prime}\right)  \frac{i}{2}\partial_{+}\right]
\end{align}
with
\begin{equation}
z(x_+) = \frac{1}{4}\int_{-\infty}^{x_{+}
}\mathrm{d}x_{+}^{\prime}\,s_{--}\left(  x_{+}^{\prime}\right).
\end{equation}
Under the coordinate transformation $x^\mu\to x^\mu +\frac{n_-^{\mu}}{2}
\,\epsilon(x_+)$, where $\epsilon$ is an arbitrary function of $x_+$,
the soft graviton field and spinor field $\xi$ have the gauge
transformations
\begin{equation}
s_{--}\rightarrow s_{--}-2\partial_{-}\epsilon(x_{+}),
\qquad
\xi \rightarrow \xi- \frac{n_-^{\mu}}{2}
\,\epsilon(x_+)\partial_\mu\xi,
\label{eq:softGT}
\end{equation}
respectively,
and $Z_{n}$ transforms as a translation operator:
\begin{equation}
Z_{n}\left(  x_{+}\right)  \rightarrow Z_{n}\left(  x_{+}\right)  \exp\left[
-\epsilon\left(  x_{+}\right)  \frac{\partial}{\partial x_{-}}\right]  .
\end{equation}

To demonstrate the decoupling of the soft graviton from the Lagrangian
we need the identity
\begin{equation}
\partial_{-}-\frac{1}{4}s_{--}\left(  x_{+}\right)  \,\partial_{+}%
=Z_{n}\left(  x_{+}\right)  \partial_{-}Z_{n}^{-1}\left(  x_{+}\right)  .
\end{equation}
We can then express the soft-collinear action
(\ref{eq:softLag}) as
\begin{align}
\int\mathrm{d}^{4}x\,{\cal L}_{c+s}^{\left(  0\right)  } &  =\int\mathrm{d}
^{4}x\,\bar{\xi}\,\frac{\not n_{+}}{2}Z_{n}\left(  x_{+}\right)  \left(
i\overrightarrow{\partial}_{-}+i\overrightarrow{\partial}_{\perp}\frac
{1}{i\overrightarrow{\partial}_{+}}i\overrightarrow{\partial}_{\perp}\right)
Z_{n}^{-1}\left(  x_{+}\right)  \xi
\nonumber \\
&  =\int\mathrm{d}^{4}x\,\bar{\xi}^{\prime}\,\frac{\not n_{+}}{2}\left(
i\overrightarrow{\partial}_{-}+i\overrightarrow{\partial}_{\perp}\frac
{1}{i\overrightarrow{\partial}_{+}}i\overrightarrow{\partial}_{\perp}\right)
\xi^{\prime},
\end{align}
where the primed fields, which are invariant under the transformation
(\ref{eq:softGT}), are defined by
\begin{equation}
\bar{\xi}^{\prime}=Z_{n}^{-1}\left(  x_{+}\right)  \bar{\xi}Z_{n}\left(
x_{+}\right)  ,\qquad\xi^{\prime}=Z_{n}^{-1}\left(  x_{+}\right)  \xi
\,Z_{n}\left(  x_{+}\right)  .\label{eq:Zdress}%
\end{equation}
Thus, a field redefinition similar to the collinear redefinition
(\ref{eq:Fredef}) eliminates the soft gravitational field from the
Lagrangian. But in this case a scattering amplitude generated by
some source operator acquires soft Wilson lines corresponding to
different light-like directions $n_{-}$, which, contrary to the collinear
Wilson lines, do not cancel.

To be specific, consider a process whose initial and final states
are clusters of highly energetic collinear particles and soft
particles $X$ including gravitons. The frame is
fixed by the time-like 4-vector
$n_{0}=(1,\vec{0}\,)  $ so that $P_{ini}^{\mu}=\left(  P_{ini}\cdot
n_{0}\right)  \,n_{0}^{\mu}$, where $P_{ini}$ is the total momentum of the
initial state. The total momentum of the $i$-th cluster $P_{i}=\left(
E_{i},\mathbf{P}_{i}\right)  $ is assumed to satisfy $E_{i}^{2}\gg P_{i}^{2}$.
We choose the reference directions as $n_{i\mp}^\mu=\left(  1,\pm\mathbf{P}%
_{i}/\left\vert \mathbf{P}_{i}\right\vert \right)  $. If we introduce the
hybrid representation~\cite{Bauer2001,Bauer2002},
e.g., $\xi\left(  x_{+},x_{-},x_{\perp}\right)  =\exp\left(  -ip_{+}%
x_{-}/2\right)  \xi_{p_{+}}\left(  x_{+},0,x_{\perp}\right)  $, then the
field redefinition with the Wilson-line operators (\ref{eq:Zdress})
results in the following factorization:%
\begin{equation}
\xi_{p_{+}}=\mathcal{Z}\left(  p_{+},x_{+}\right)  \xi_{p_{+}}^{\prime}
,\qquad\mathcal{Z}\left(  p_{+},x_{+}\right)  =\exp\left[  -\frac{i}{4}%
\int_{-\infty}^{x_{+}}\mathrm{d}x_{+}^{\prime}s_{--}\left(  x_{+}^{\prime
}\right)  \frac{1}{2}\,p_{+}\right]  .
\end{equation}
Therefore, the soft graviton interaction with the $i$-th collinear cluster
contributes the factor $\mathcal{Z}\left(  n_{i+}\cdot
P_{i},x_{+}\right)  $ to the amplitude.
Using translation invariance to move the argument $x_+$ of
$\mathcal{Z}\left(  p_{+},x_{+}\right)$ to 0, and momentum
conservation, the total amplitude $M\left(  P_{1},\ldots P_{N}\right)
$, where $N$ is a number of clusters, thus factorizes into the product $M\left(
P_{1},\ldots P_{N}\right)  =S_{N}\,\tilde{M}\left(  P_{1},\ldots P_{N}\right)
$, where $\tilde{M}$ is independent of the soft graviton field and $S_{N}$ is
the soft factor:
\begin{equation}
S_{N}=\left\langle X\left\vert \prod_{i=1}^{N}\mathcal{Z}\left(  n_{i+}\cdot
P_{i},0\right)  \right\vert 0\right\rangle . \label{eq:SoftFactor}
\end{equation}
This is precisely the eikonal form of the amplitude which has been
established by Weinberg~\cite{Weinberg1965} and which is studied in detail
in Refs.~\cite{Naculich2011, White2011}.

\subsection*{6 Conclusion}

In a gauge theory with a massless vector boson as the interaction carrier,
large logarithmic corrections appear in the collinear limit when particle
energies are much larger than their masses and splitting angles tend to
zero. The interplay with soft singularities results in double
logarithmic corrections and the associated non-trivial dynamics
has been intensively studied for collider physics applications. Much
recent progress in this area of strong and electroweak physics is based
on the construction of an effective theory with explicit separation
of soft and collinear modes --
SCET~\cite{Bauer2001,Bauer2002,Beneke2002,Beneke2003}.
Motivated by this and recent work on collinear gravitational
interactions~\cite{Akhoury2011}, we discussed in this paper
the effective field theory description of soft and collinear
gravitons. It is worth noting that
the effective Lagrangians derived here apply
to graviton interactions at Planckian and even trans-Planckian
energies $E$ provided the transverse momentum scale $p_\perp\sim
\lambda E$ is smaller than the Planck scale.

Soft-collinear gravity is quite different from SCET as far as
collinear interactions are concerned. In the effective Lagrangian,
in leading power in the expansion in $p_\perp/Q$, the collinear sector
is trivial and only eikonal-type soft graviton interactions appear.
The spin of the graviton prohibits singular collinear splittings
like \textit{particle }$\rightarrow$\textit{ particle +
graviton}, thus there is no proliferation of graviton radiation
collinear to a very energetic particle, unlike the case of gauge
boson radiation.

However, the effective theory contains the collinear metric field
$h_{++}$ with an unusual negative scaling power $1/\lambda$,
which complicates the discussion.
The coupling to a non-collinear environment potentially contributes through
real or virtual radiation which interferes with purely collinear splittings.
The suppression of collinear self-interactions can be lifted, if
the interference occurs through the field with negative
scaling power. Nevertheless, contrary to gauge theories with massless
vector bosons, collinear divergences, while present in
individual diagrams, cancel in physical processes. In the effective
field theory treatment this can be shown in various ways.
The interference with the non-collinear environment occurs
in leading power only through an unphysical metric component,
$h_{++}(x_-)$, which can be eliminated in light-cone gauge. In
covariant gauge, the coupling of $h_{++}$ can be removed by
going to a different coordinate frame, or by the
universal field redefinition (\ref{eq:Fredef}), which in
turn is closely related to the coordinate and gauge transformation.
Our result complements the diagrammatic proof~\cite{Akhoury2011}
and extends it to a general covariant gauge. The final
cancellation of collinear singularities then occurs upon using
translation invariance, i.e. energy-momentum conservation, as
was already observed in Ref.~\cite{Weinberg1965} in the
soft limit.

\vspace*{0.5em}
\noindent
\subsubsection*{Acknowledgement}
We thank Slava Rychkov for asking questions that triggered our interest
in this subject. This work is supported in part by the Gottfried Wilhelm
Leibniz programme of the Deutsche Forschungsgemeinschaft (DFG).

\bibliographystyle{elsarticle-num}

\end{document}